\documentstyle[epsfig,12pt]{article}
\begin{document}
\onecolumn
\begin{titlepage}
\begin{center}
{\Large \bf Exact Relativistic Two-Body Motion\\
in Lineal Gravity} \\ \vspace{2cm}
R.B. Mann \footnotemark\footnotetext{email: 
mann@avatar.uwaterloo.ca} and D. Robbins\footnotemark\footnotetext{email: 
dgr@gpu.srv.ualberta.ca}
\\
\vspace{0.5cm}
Dept. of Physics,
University of Waterloo
Waterloo, ONT N2L 3G1, Canada\\
and \\
\vspace{1cm}
T. Ohta \footnotemark\footnotetext{email:
t-oo1@ipc.miyakyo-u.ac.jp}\\
\vspace{0.5cm} 
Department of Physics, Miyagi University of Education,
Aoba-Aramaki, Sendai 980, Japan\\
\vspace{1.5cm}
\vspace{1.5cm}
\today\\
\end{center}
\begin{abstract}
The N-body problem for one-dimensional self-gravitating systems has 
been often studied to test theories of galactic evolution
and statistical mechanics. We consider the general relativistic version
of this system and obtain the first exact solution to the 2-body problem 
in which spacetime is not flat. In the equal mass 
case we obtain an explicit expression for the proper separation of the two
masses as a function of their mutual proper time.
\end{abstract}
\end{titlepage}
\onecolumn

\vspace{5mm}

 $N$-body self-gravitating systems have a long history in physics
and are of interest in studying both star systems (small $N\geq 2$)
and galactic evolution (large $N$). One-dimensional (or lineal) 
models of such systems have been of particular interest \cite{yawn}
in that they avoid some difficulties due to three dimensions, including
evaporation, singularities, and energy dissipation in the form of 
gravitational radiation, as well as admitting a level of
computational and analytic analysis which is dramatically simpler.

We consider in this paper the $N$-body problem for a relativistic
self-gravitating lineal system and formulate a canonical theory determining its
Hamiltonian. In the 2-body
case we obtain an exact solution (valid to all orders
in the gravitational coupling $\kappa$) in which spacetime outside the
moving matter sources is not flat. In the equal mass case we
obtain an explicit expression for the proper separation of the
two point masses as a function of their mutual proper time.
This is the first non-perturbative relativistic curved-spacetime treatment
of this problem, providing new avenues for investigation 
of one-dimensional self-gravitating systems.

For our lineal self-gravitating system we choose a modification of
Jackiw-Teitelboim lineal gravity, in 
which the scalar curvature is equated to a cosmological constant
\begin{equation}\label{RL}
R - \Lambda = 0
\end{equation}
in the absence of other matter fields \cite{JT}. This model has
been of considerable interest as a model theory of quantum
gravity \cite{jtrfs}. Here we couple $N$ point masses to this theory 
so that we have a generally covariant self-gravitating system
with non-zero curvature outside the point sources. We do not include
collisional terms, so that the bodies pass through each other.

Since the Einstein action is a topological invariant in 2 spacetime
dimensions, we must incorporate a scalar (dilaton) field into our
action, which we take to be
\begin{eqnarray}\label{act1}
I&=&\int d^{2}x\left[
\frac{1}{2\kappa}\sqrt{-g}
\left\{\Psi R + \frac{1}{2}g^{\mu\nu}\nabla_{\mu}\Psi\nabla_{\nu}\Psi
 + \Lambda \right\}\right] + I_P
\end{eqnarray}
where $\Psi$ is the  dilaton field, $g_{\mu\nu}$ and
$g$ are the metric and its determinant and $R$ is the Ricci
scalar,  
with $\kappa=8\pi G/c^4$.  $I_P$ is the action of $N$ point masses
minimally coupled to gravity
\begin{eqnarray}\label{act1a}
I_P&=& -\int d^{2}x\left[\sum_{a=1}^N m_{a}\int d\tau_{a}
\left\{-g_{\mu\nu}(x)\frac{dz^{\mu}_{a}}{d\tau_{a}}
\frac{dz^{\nu}_{a}}{d\tau_{a}}\right\}^{1/2}\delta^{2}(x-z_{a}(\tau_{a}))
\right] 
\nonumber
\end{eqnarray}
\noindent
where $\tau_{a}$  is the proper time of $a$-th particle.
Variation of the 
action (\ref{act1}) with respect to the metric, dilaton field, 
and particle coordinates yields the field equations
\begin{equation}\label{RTgeo}
R-\Lambda=\kappa T^{P\mu}_{\;\;\mu} \qquad
\frac{d}{d\tau_{a}}
\left\{\frac{dz^{\nu}_{a}}{d\tau_{a}}\right\}
+\Gamma^\nu_{\alpha\beta}(z_a)
\frac{dz^{\alpha}_{a}}{d\tau_{a}}
\frac{dz^{\beta}_{a}}{d\tau_{a}}=0 
\end{equation}
\begin{equation}\label{e4}
\frac{1}{2}\nabla_{\mu}\Psi\nabla_{\nu}\Psi
-g_{\mu\nu}\left(\frac{1}{4}\nabla^{\lambda}\Psi\nabla_{\lambda}\Psi
-\nabla^{2}\Psi\right)
-\nabla_{\mu}\nabla_{\nu}\Psi=\kappa T^P_{\mu\nu} 
+ \frac{\Lambda}{2}g_{\mu\nu}\end{equation}
where the stress-energy due to the point masses is
\begin{equation}
T^P_{\mu\nu} = \sum_{a=1}^N m_{a}\int d\tau_{a}\frac{1}{\sqrt{-g}}
g_{\mu\sigma}g_{\nu\rho}\frac{dz^{\sigma}_{a}}{d\tau_{a}}
\frac{dz^{\rho}_{a}}{d\tau_{a}}\delta^{2}(x-z_{a}(\tau_{a})) 
\nonumber
\end{equation}
and is conserved.  We observe that (\ref{RTgeo}) is a closed system of 
$N+1$ equations for which one can solve for the single metric degree
of freedom and the $N$ degrees of freedom of the point masses; it
reduces to (\ref{RL}) if all masses vanish. The
evolution of the dilaton field is governed by the evolution of the
point-masses via (\ref{e4}). The left-hand side of (\ref{e4})
is divergenceless (consistent with the conservation
of $T_{\mu\nu}$), yielding only one independent equation to determine
the single degree of freedom of the dilaton.

Working in the canonical formalism we make use of the decomposition 
$\sqrt{-g}R=-2\partial_{0}
(\sqrt{\gamma}K)+2\partial_{1}
(\sqrt{\gamma}N^{1}K-\gamma^{-1}\partial_{1}N_{0})$ 
where the extrinsic curvature
$K=(2N_{0}\gamma)^{-1}(2\partial_{1}N_{1}
-\gamma^{-1}N_{1}\partial_{1}\gamma -\partial_{0}\gamma)$,
and rewrite  the action (\ref{act1}) in the form \cite{ohtarobb}
\begin{equation}\label{e9}
I=\int dx^{2}\left\{\sum_{a}p_{a}\dot{z}_{a}\delta(x-z_{a}(x^{0}))
+\pi\dot{\gamma}+\Pi\dot{\Psi}+N_{0}R^{0}+N_{1}R^{1}\right\} 
\end{equation}
where  $\gamma=g_{11},  N_{0}= (-g^{00})^{-1/2}, N_{1}= g_{10}$,
$\pi$ and $\Pi$ are conjugate momenta to $\gamma$ and $\Psi$
respectively. The quantities $N_0$ and $N_1$ are Lagrange multipliers 
which enforce the constraints $R^0 = 0 = R^1$, where
\begin{eqnarray}
R^{0}&=&-\kappa\sqrt{\gamma}\gamma\pi^{2}+2\kappa\sqrt{\gamma}\pi\Pi
+\frac{(\Psi^{\prime})^{2}}{4\kappa\sqrt{\gamma}}-
\left(\frac{\Psi^{\prime}}{\kappa\sqrt{\gamma}}\right)^{\prime}
+\frac{\Lambda}{2\kappa}\sqrt{\gamma}
-\sum_{a}\sqrt{\frac{p^{2}_{a}}{\gamma}+m^{2}_{a}}\;
\delta(x-z_{a}(x^{0}))
\nonumber \\
R^{1}&=&\frac{\gamma^{\prime}}{\gamma}\pi-\frac{1}{\gamma}\Pi\Psi^{\prime}
+2\pi^{\prime}
+\sum_{a}\frac{p_{a}}{\gamma}\delta(x-z_{a}(x^{0}))
\end{eqnarray}
with the symbols $(\;\dot{}\;)$ and  $(\;^{\prime}\;)$
denoting $\partial_{0}$ and $\partial_{1}$, respectively.
We identify the dynamical and gauge ({\it i.e.} co-ordinate) 
degrees of freedom by writing the generator arising from the variation of the 
action at the boundaries in terms of $(\Psi^{\prime}/\sqrt{\gamma})^{\prime}$ 
and $\pi^{\prime}$, which we can easily solve for since 
these are the only linear terms in the
constraints. We then  fix the frame of the physical space-time 
coordinates in a manner similar to the $(3+1)$-dimensional case \cite{adm}. 

Carrying out this procedure,
we find that we can consistently choose the coordinate conditions
$\gamma=1$ and $\Pi=0$. Eliminating the constraints, the action 
(\ref{e9}) then reduces to 
\begin{equation}\label{hamact}
I=\int d^{2}x\left\{\sum_{a}p_{a}\dot{z}_{a}\delta(x-z_{a})
-\cal H\mit\right\}\;\;.
\end{equation}
where the reduced Hamiltonian is $H=\int dx {\cal H} 
=-\frac{1}{\kappa}\int dx \triangle\Psi$ ,
where $\triangle \equiv \partial^{2}/\partial x^{2}$,
and $\Psi = \Psi(x,z_{a},p_{a})$ and is 
understood to be determined from
the constraint equations which are now
\begin{eqnarray}\label{psicon}
\triangle\Psi-\frac{(\Psi^{\prime})^{2}}{4}
+\kappa^{2}\pi^{2} 
- \frac{\Lambda}{2}+\kappa\sum_{a}\sqrt{p^{2}_{a}+m^{2}_{a}}
\delta(x-z_{a})&=&0 \nonumber \\
2\pi^{\prime}+\sum_{a}p_{a}\delta(x-z_{a})&=&0  \; .
\end{eqnarray}
When $\Lambda=0$ the Hamiltonian reduces to that considered in ref. \cite{yawn}
in the non-relativistic limit \cite{ohtarobb}.

For $N=2$ we solve these equations exactly following the method given in ref. 
\cite{rbmohtaprd}. First, for $z_{2}<z_{1}$, we divide 
spacetime into three regions: $z_{1}<x$ ((+) region), 
$z_{2}<x<z_{1}$ ((0) region) and $x<z_{2}$ ((-) region) and set
$\Psi=-4\mbox{log}|\phi|$ and $\pi=\chi^\prime$. In each region
$\chi$ is a sum of terms linear in $x$ and $\phi$ 
obeys a harmonic oscillator equation, whose solutions are linear
combinations of exponential functions of $x$ in the $\pm$ regions
and of either trigonmetric or exponential functions of $x$ in
the $0$ region depending on the size of $\Lambda$ relative to the other integration
constants.   Matching these
solutions at the boundaries $x=z_1$ and $x=z_2$ of each region
allows a determination of the coefficients of these linear
combinations in the $+$ and $-$ regions in terms of those in the $0$ region.

The magnitudes of both $\phi$ and $\chi$ increase with 
increasing $|x|$ and so we must employ a boundary condition
which guarantees that the surface terms obtained in passing from
(\ref{act1}) to (\ref{e9}) vanish and the Hamiltonian remain finite. 
A perturbative analysis indicates that this condition must be 
$\Psi^{2}-4\kappa^{2}\chi^{2}+2\Lambda x^2 = C_{\pm}x$ in 
the $\pm$ regions.  Incorporating this into the matching
conditions determines the coefficients of the linear combination
in the $0$ region in terms of the momenta and positions of the bodies.  

This procedure is sufficient to solve for $N_0$, $N_1$ and all other field
variables exactly. Repeating the analysis for $z_1 < z_2$ yields a similar
solution with $p_i\to -p_i$, and so the full solution is a function of
$(z_i,\tilde{p}_i)$, where $\tilde{p}_i = p_i \mbox{sgn}(z_1-z_2)$.
The resultant expressions are rather long and cumbersome.
However it is straightforward to show that the canonical equations imply
conservation of the total momenta $p_1 + p_2$. Choosing a center of inertia 
frame with $p_1=-p_2=p$,  yields considerable simplification of the exact
solution,  which we write employing the notation
$K=\sqrt{ (\kappa X)^{2}-\frac{\Lambda}{2}}$, 
$K_0 = \sqrt{Y_0^{2}-\frac{\Lambda}{2}}$ and
\begin{eqnarray}\label{nota}
K_{1,2} \equiv  2K_{0}+2K -\kappa\sqrt{p^{2}+m_{2,1}^{2}}
&\qquad&
\hat{K} \equiv K_{0}-K+\frac{\kappa\epsilon}{2}\tilde{p}
\nonumber \\
{\cal M}_{1,2}\equiv \kappa\sqrt{p^{2}+m_{1,2}^{2}}+2K_{0}-2K
&\qquad&
Y_{0} \equiv \kappa\left[X-\frac{\epsilon}{2}\tilde{p}\right] \quad .
\nonumber 
\end{eqnarray}
We obtain 
\begin{equation}\label{phi-sol1}
\phi_{\pm}(x)=A_{\pm}e^{\pm\frac{1}{2}K x} \quad
\phi_{0}(x)=A_{0}e^{\frac{x}{2}K_{0}}+B_{0}e^{-\frac{x}{2}K_{0}} 
\end{equation}
in the $(+)$, $(-)$ and $(0)$ regions respectively, where
\begin{eqnarray}\label{ab+-0}
A_{\pm}&=&\left(\frac{K_{1,2}}{{\cal M}_{1,2}}\right)^{1/2}e^{-\frac{\hat{K}}{4}
(z_{1}-z_{2}) \mp \frac{1}{2}K z_{1,2}}
\\
\{A_0,B_0\}&=&\frac{(K_{2,1}{\cal M}_{2,1})^{1/2}}{4K_{0}}e^{-\frac{\hat{K}}{4}
(z_{1}-z_{2})\mp\frac{1}{2}K_{0}z_{2,1}} \quad .
\nonumber 
\end{eqnarray}
The metric components are
\begin{eqnarray}\label{N10}
N_{0(\pm)}&=& A\phi^{2}_{\pm} \qquad N_{0(0)}= A\phi^{2}_{(0)}\nonumber \\
N_{1(\pm)}&=&\pm\epsilon \frac{\kappa X}{K} \left(A\phi^{2}_{\pm}-1\right)\\
N_{1(0)}&=&\epsilon \left\{A{Y_{0}}\left[2A_{0}B_{0}x
+\frac{A_{0}^{2}\;e^{K_{0}x}
-B_{0}^{2}\;e^{-K_{0}x}}{K_{0}}\right]+D_{0}\right\}\nonumber
\end{eqnarray}
where 
%
$A=\frac{16\kappa K_{0} X }{J_+ K}\;e^{\frac{1}{2}\hat{K}(z_{1}-z_{2})}$,
$D_0 = \kappa\frac{J_{-} X}{J_+ K}$
and
\begin{equation}\label{J}
J_\pm = 2\left(\frac{Y_{0}}{K_{0}} + \frac{\kappa X}{K}\right)\left(K_{1}
\pm K_{2}\right) 
-2K_{1}K_{2}
\left(\frac{Y_{0}}{K_{0}}-\frac{\kappa X}{K}\right)
\left(\frac{1}{{\cal M}_{1}} \pm \frac{1}{{\cal M}_{2}}\right)
-\frac{Y_{0}}{K_{0}}K_{1}K_{2}(z_{1}\mp z_{2}) .
\end{equation}

The solution for $z_1 < z_2$ is obtained by interchanging 
the suffices 1 and 2 in the preceding solution.

The matching conditions at $(z_1,z_2)$ force the relation
\begin{equation}\label{eqX}
K_{1}K_{2}={\cal M}_{1}{\cal M}_{2}e^{K_{0}|z_{1}-z_{2}|}
\end{equation}
which determines $X$. From this the Hamiltonian
\begin{equation}\label{Ham1}
H = -\frac{1}{\kappa}\int dx\triangle\Psi
= -\frac{1}{\kappa}\left[\Psi^{\prime}\right]^{\infty}_{-\infty}
= \frac{4K}{\kappa}\;\;.
\end{equation}
may be explicitly determined as a function of the coordinates and
momenta of the particles. 

The parameter $\epsilon=\pm 1$, and is a constant of integration associated
with the  metric degree of freedom.  Under time reversal, solutions with
$\epsilon=1$ transform into those with $\epsilon=-1$, ensuring invariance of the whole 
theory under this symmetry. 
It is straightforward to show from this solution
that the Ricci scalar is equal to the cosmological constant everywhere
except at the locations  $(z_1(t),z_2(t))$ of the point masses.  
Hamilton's equations imply that 
\begin{eqnarray}\label{p1}
\dot{p} &=&-\frac{4 X}{K} \frac{K_{0}K_{1}K_{2}}{J_{+}}\\
\dot{z}_{i}
&=& (-1)^{i+1}\frac{\kappa X}{K} \left(\epsilon+\frac{16}{J_{+}}\frac{K_{0}K_{i}}{{\cal M}_{i}}
\left\{\frac{p}{\sqrt{p^{2}+m_{i}^{2}}}-\epsilon\frac{\kappa X}{K}\right\}\right)
\nonumber
\end{eqnarray}
are the dynamical equations for the 2-body system coupled to gravity, where $i=1,2$.

The defining equation for the Hamiltonian is, using eqs. (\ref{eqX}) and (\ref{Ham1}),
\begin{equation}\label{defineH}
\tanh(\frac{\kappa{\cal J}}{8}|r|) = \frac{{\cal J}(B_1+B_2)}{{\cal J}^2+B_1 B_2}
\end{equation} 
where $B_{1,2} = H-2\sqrt{p^2+m_{1,2}^2}$,
${\cal J}^2=(\sqrt{H^2+8\Lambda/\kappa^2}-2\epsilon\tilde{p})^2-8\Lambda/\kappa^2$
and $r\equiv z_{1}-z_{2}$. 
For a given $\Lambda \geq -(\kappa H)^2/8$,
equation (\ref{defineH}) describes the surface in $(r,p,H)$ space of all
allowed phase-space trajectories. Since  $H$ is a constant of the motion, 
(a fact easily verified by differentiation of (\ref{defineH}) 
with respect to $t$) a given trajectory in the $(r,p)$ plane
is uniquely determined by setting  $H=H_0$ in (\ref{defineH}). 

In the equal mass case the canonical equations of motion (\ref{p1})
can be solved exactly.  The proper time of each particle is
\begin{equation}
d\tau=dt\;N_{0}(z_{a})\frac{m}{\sqrt{p^2+m^2}}
=dt\;\frac{16\kappa K_{0}K_{1}X}{J_{+}K{\cal M}_{1}}\frac{m}{\sqrt{p^2+m^2}}
\nonumber
\end{equation}
yielding upon insertion in (\ref{p1}) the solution 
$p = \frac{\epsilon m}{2}(f(\tau)-1/f(\tau))$, where
\begin{equation}
f(\tau) = \cases{
\frac{\frac{H}{m}(1+ \sqrt{\gamma_H})}
{1+\sqrt{-\gamma_m}
 \frac{\sigma+\frac{m^2}{H}\sqrt{-\gamma_m}
   \tan\left[\frac{\epsilon\kappa m}{8}\sqrt{-\gamma_m}(\tau-\tau_{0})\right]}
{\frac{m^2}{H}\sqrt{-\gamma_m}
-\sigma \tan\left[\frac{\epsilon\kappa m}{8}\sqrt{-\gamma_m}(\tau-\tau_{0})\right]}}
&$\gamma_m < 0 $
\cr
\frac{1+\sqrt{\gamma_H}}
{\frac{m}{H}+\frac{\sigma}{m-\sigma\frac{\epsilon\kappa H}{8}(\tau-\tau_{0})}}
&$\gamma_m=0 $
\cr
\frac{
\frac{H}{m}\left(1+\sqrt{\gamma_H}\right)
\left\{1-\eta\;e^{\frac{\epsilon\kappa m}{4}\sqrt{\gamma_m}(\tau-\tau_{0})}\right\}}
{1+\sqrt{\gamma_m}
+\left(\sqrt{\gamma_m}-1\right)
\;\eta\;e^{\frac{\epsilon\kappa m}{4}\sqrt{\gamma_m}(\tau-\tau_{0})}}
&$\gamma_m>0$ 
}
\end{equation}
where $\gamma_x\equiv 1+ \frac{8\Lambda}{\kappa^2 x^2}$,
$\sigma = (1+\sqrt{\gamma_H})
(\sqrt{p_{0}^2+m^2}-\epsilon p_{0})-\frac{m^2}{H}$, and
$\eta=\frac{\sigma -\frac{m^2}{H}\sqrt{\gamma_m}}
{\sigma + \frac{m^2}{H}\sqrt{\gamma_m}}$,
with $p_0$ the initial momentum at $\tau=\tau_0$.
It is then straightforward to obtain an exact expression for $r$ 
as a function of $\tau$, either directly from the second of
eqs. (\ref{p1}) or by inserting the expression
for $p(\tau)$ into (\ref{defineH}) and solving for $r$.

Analysis of these solutions shows that two types of motion are possible,
depending on the amount of energy and the value of the cosmological constant.
For $\Lambda < \frac{\kappa^2m^4}{2\left(H^2-4m^2\right)}\equiv \Lambda_c$ 
the two particles will always execute periodic motion, with period
\begin{equation}
T=\cases{\frac{16}{\sqrt{\kappa^2m^2+8\Lambda}}\tanh^{-1}\left(\frac{
\sqrt{\kappa^2m^2+8\Lambda}\sqrt{H^2-4m^2}}{\kappa Hm}\right)
&$\gamma_m>0$ \cr
\frac{16\sqrt{H^2-4m^2}}{\kappa Hm}&$\gamma_m=0$ \cr
\frac{16}{\sqrt{-\kappa^2m^2-8\Lambda}}\tan^{-1}\left(\frac{
\sqrt{-\kappa^2m^2-8\Lambda}\sqrt{H^2-4m^2}}{\kappa Hm}\right)
&$\gamma_m<0$}
\end{equation}
which can be  obtained from the initial and final values of $p$ when $r=0$ from 
(\ref{defineH}). If $\Lambda < 0$ this is the only state of motion possible.
However if $\Lambda>0$ there also exist a countably infinite set of 
unbound states of motion, in which the particles begin with some (sufficiently large)
separation, approach one another at some minimal value of $|r|$, and then
move apart from one another toward infinite separation.

For $\Lambda > \Lambda_c$, only unbounded motion is possible. If the
initial momentum is sufficiently small, the particles will cross one another 
before receding toward infinity.  Otherwise they simply 
approach one another at some minimal value of $|r|$ and then reverse direction
toward infinity as previously mentioned.  One peculiar feature in this regime
is that the two particles diverge to infinite separation at finite $\tau$!  
Specifically, if the above condition
is satisfied, then the determining equation requires that $r\to\infty$ as
 $p$ approaches the value
$\frac{\epsilon}{2\kappa}\left(\sqrt{\kappa^2H^2+8\Lambda}
+\sqrt{8\Lambda}\right)$.  
The corresponding value of $\tau$ at which this will occur can be found from
the $\gamma_m > 0$
solution above.  
The explicit expression for this $\tau$ is cumbersome and is omitted.

In figure 1 we plot $r$ vs. $\tau$ for two bodies initially at $r=0$
for fixed $\Lambda=-1.5$ and $H=16$ for 
several different values of $m$. As the motion becomes more relativistic
(i.e. $m$ gets smaller) we find that a second maximum develops in the curve,
which then vanishes for very small $m$. We find this behaviour for a broad range
of values of $\Lambda<0$ and $H$.
\begin{figure}
\begin{center}
\epsfig{file=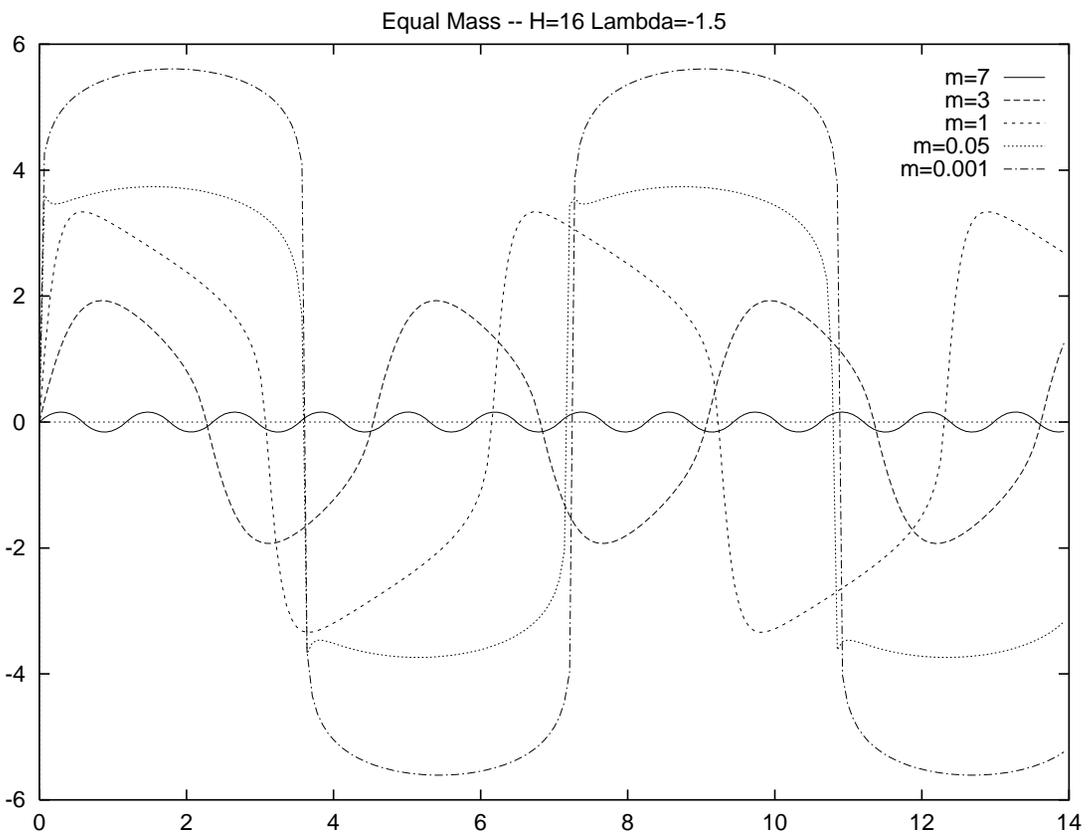,width=0.8\linewidth}
\end{center}
\caption{A sequence of equal mass curves for $\Lambda=-1.5$, $H=16$.
Note the presence of the second maximum for $m=0.05$.}
\label{fig1}
\end{figure}
\begin{figure}
\begin{center}
\epsfig{file=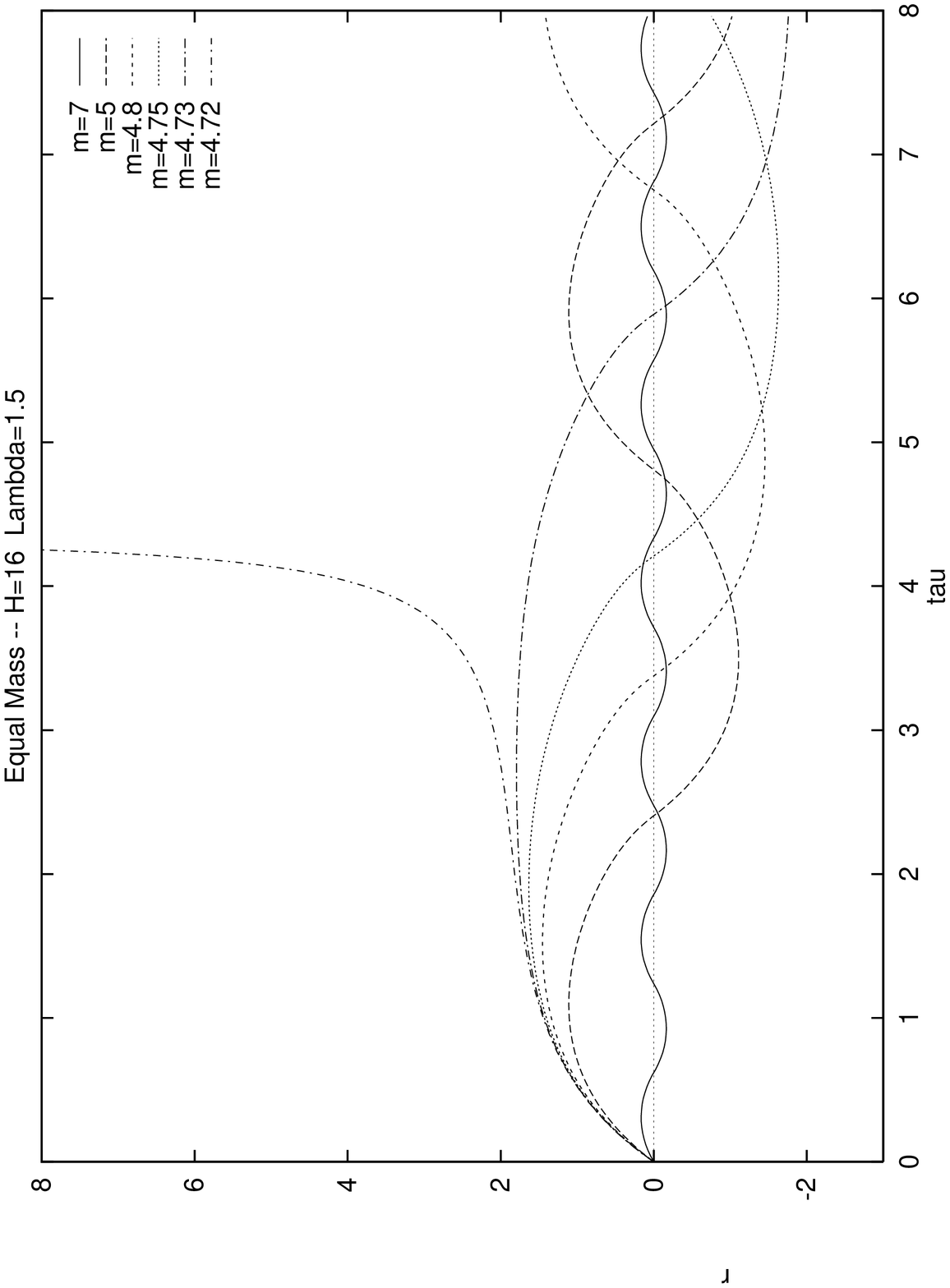,width=0.8\linewidth}
\end{center}
\caption{A sequence of equal mass curves for $\Lambda=1.5$, $H=16$.
The motion becomes unbounded between $m=4.72$ and $m=4.73$.}
\label{fig2}
\end{figure}
Figure 2 contains an analogous plot for $\Lambda =1.5$, showing the transition
from bounded to unbounded motion.  As $\Lambda\to\Lambda_c$ (i.e. as $m$ becomes small)
the particles rapidly separate, remaining nearly stationary for an increasingly
large period of proper time before coming together again. At $\Lambda=\Lambda_c$
this separation time becomes infinite, and for $\Lambda > \Lambda_c$, the
separation diverges at finite $\tau$.

To summarize, we have obtained an exact solution to the 2-body problem in
relativistic lineal gravity in which spacetime is not flat, in contrast to
other lower-dimensional solutions to this problem in which spacetime
is either flat outside matter \cite{rbmohtaprd,Bellini,Gottmov} or the matter sources 
are stationary \cite{DJT}.  The system is described by a conservative Hamiltonian 
in the canonical formalism. Many interesting features of this problem remain 
to be explored, including an extension to the unequal-mass case, the transition 
from bounded to unbounded motion for $\Lambda > 0$, the appearance/disappearance of 
a second maximum for $\Lambda <0$, and an investigation of the statistical properties
of the model for large $N$.  In this last case many-body forces play an important 
role -- as shown in ref. \cite{ohtarobb},
in the post-Newtonian approximation $n$-body forces appear at order $\kappa^{n-1}$.
We intend to make these the subjects of future investigation.

This work was supported in part by the
Natural Sciences and Engineering Research Council of Canada.

\end{document}